\documentclass[smus]{snow2e}
\usepackage{graphics}

\begin{document}
\begin{titlepage}
\hbox to \hsize{
\hfill\vtop{\hbox{}
\vspace{2.cm}
\hbox{\large Fermilab-Conf-96/422-T}
\vspace{2.mm}
\hbox{\large November 1996} } }
\vspace{1.1cm}
\begin{center}
{\Large \sc Measuring the mass of the $W$ at the LHC
\footnote{To appear in the {\it Proceedings of the 1996 DPF/DPB Summer Study on New Directions for High-Energy Physics (Snowmass 96)}.
}\\[1.cm]}
{\large \sc S.~Keller and J.~Womersley\\[2.mm]}
{\large \it Fermilab, P.O. Box 500, Batavia, IL 60510, USA\\[0.5cm]}
\vspace{1.cm}

{\large \bf
ABSTRACT \\[6.mm]}
\end{center}

{ \baselineskip=18pt \large
We explore the ability of the Large Hadron Collider to measure the mass of 
the $W$ boson.
We believe that a precision better than $\sim 15$~MeV 
could be attained, based on a year of operation 
at low luminosity ($10^{33}\,{\rm cm}^{-2}\,{\rm s}^{-1}$).
If this is true, this measurement will be
the world's best determination of the $W$ mass. We feel
this interesting opportunity warrants investigation in more detail.
}

\end{titlepage}

\title{Measuring the mass of the $W$ at the LHC}

\author{S.~Keller and J.~Womersley\\ {\textit{Fermi National Accelerator
Laboratory, Batavia, IL 60510\thanks{operated by the Universities
Research Association, Inc., for 
the U.S. Department of Energy}
}}}

\maketitle

\thispagestyle{empty}\pagestyle{empty}

\begin{abstract} 
We explore the ability of the Large Hadron
Collider to measure the mass of the $W$ boson.
We believe that a precision better than $\sim 15$~MeV 
could be attained, based on a year of operation 
at low luminosity ($10^{33}\,{\rm cm}^{-2}\,{\rm s}^{-1}$).
If this is true, this measurement will be
the world's best determination of the $W$ mass. We feel
this interesting opportunity warrants investigation in more detail.
\end{abstract}
\vskip 1cm
The mass of the $W$ boson, $m_W$, is one of
the fundamental parameters of the Standard Model.  
As is well known~\cite{wolf}, a precise measurement
of $m_W$, along with other precision electroweak measurements,  
will lead, within the Standard Model, 
to a strong indirect constraint on the mass of 
the Higgs boson.  The precise measurement of 
$m_W$ is therefore a priority
of future colliders.  LEP2 and Run II at Fermilab  
($\int\!{\cal L}dt=1$~fb$^{-1}$)
are aiming for an 
uncertainty on $m_W$ of about 40 MeV~\cite{lep} and 
35 MeV~\cite{tev2000}, respectively. 
A recent study~\cite{tevah} has shown that an upgrade of the Tevatron, 
beyond Run II, might be possible,
with a goal of an overall integrated luminosity 
of ${\cal O}(30 \mbox{fb}^{-1})$
and a precision on $m_W$ of about 15 MeV. 
In this short contribution we take a very first look at the potential 
to measure $m_W$ at the Large Hadron Collider (LHC). 

It has often been claimed that LHC detectors will not be able to trigger
on lepton with sufficiently low transverse momentum ($p_T$)
to record the $W$ sample
needed for a measurement of $m_W$.  While this may be true at 
the full LHC luminosity ($10^{34}\,{\rm cm}^{-2}\,{\rm s}^{-1}$)
it does not appear to be the case at $10^{33}\,{\rm cm}^{-2}\,{\rm s}^{-1}$.  
Based on a full GEANT simulation of the calorimeter, 
the CMS isolated electron/photon trigger~\cite{cmsetrig}  
should provide an acceptable rate ($<5$~kHz at level 1) for a threshold
setting of $p_T^{e,\gamma} > 15$~GeV/c. This trigger will be
fully efficient for
electrons with $p_T^{e} > 20$~GeV/c.
The CMS muon trigger~\cite{cmsmutrig} should also operate
acceptably with a threshold of $p_T^{\mu} > 15-20$~GeV/c
at $10^{33}\,{\rm cm}^{-2}\,{\rm s}^{-1}$.
It is likely that the accelerator will operate for at least a year 
at this `low' luminosity to allow
for studies which require heavy quark 
tagging (e.g., $B$-physics).
This should provide an integrated luminosity of the order of $10fb^{-1}$
and therefore an ample dataset for a measurement of $m_W$.

The mean number of interactions per crossing, $I_C$, is about 2
at the low luminosity.
This is actually lower than the number of interactions
per crossing during
the most recent run (IB) at the Fermilab Tevatron.  In this 
relatively quiet environment it should be straightforward to
reconstruct electron and muon tracks with good efficiency.
Furthermore, both the ATLAS and CMS detectors offer advances over their 
counterparts at the Tevatron for lepton identification and measurement:
they have 
precision electromagnetic calorimetry (liquid argon and PbWO$_4$ crystals, 
respectively) and precision muon measurement (air core toroids and high
field solenoid, respectively). 

The missing transverse energy will also be well
measured thanks to the small number of interactions per crossing
and the large pseudorapidity coverage ($|\eta|<5$) of the
detectors.  The so-far standard transverse-mass technique for determining
$m_W$ should thus continue to be applicable.    
This is to be contrasted with the problem that the increase 
in $I_C$ will create for Run II at the Tevatron. 
In Ref.~\cite{tev2000}, it was shown that it will substantially degrade the 
measurement of the missing transverse energy and therefore 
the measurement of $m_W$.   

\begin{figure}[t]
\leavevmode
\begin{center}
\resizebox{!}{17cm}{
\includegraphics{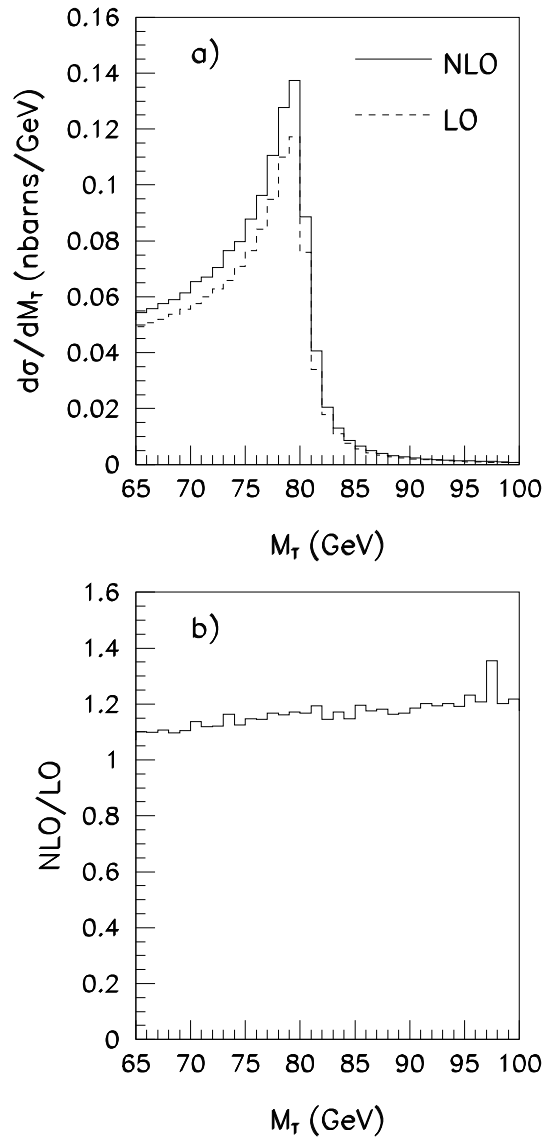}}
\end{center}
\caption{a) LO calculation (dashed line) and 
NLO QCD calculation (solid line) of the $m_T$ distribution at the LHC.  
See text for the cuts.  
b) ratio of the NLO calculation over the LO calculation
as a function of $m_T$.}
\label{fig:lhc}
\end{figure}

\begin{figure}[t]
\leavevmode
\begin{center}
\resizebox{!}{17cm}{
\includegraphics{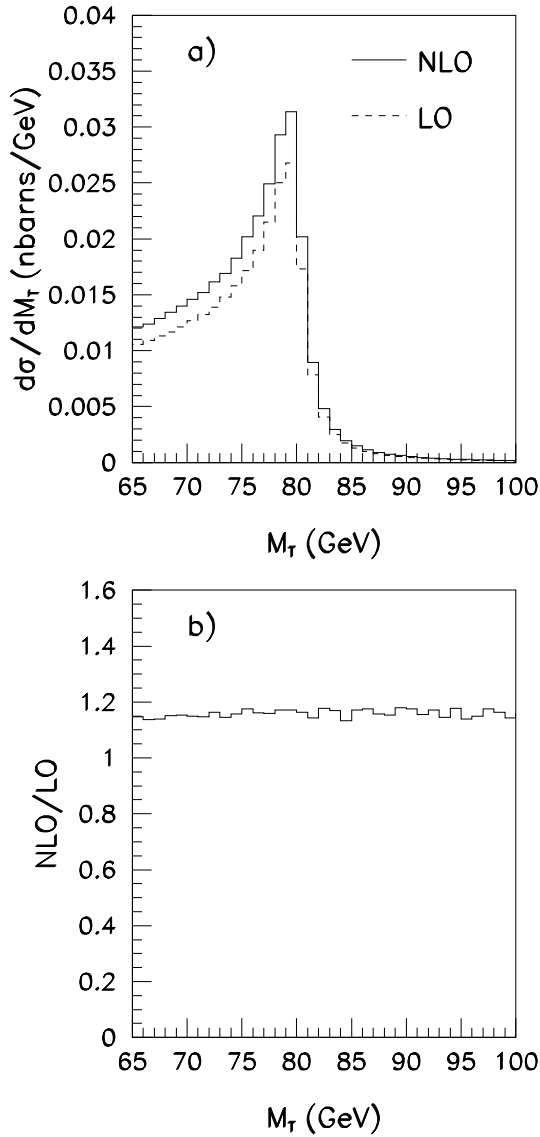} }
\end{center}
\caption{Same as in Fig.~\ref{fig:lhc} but for the Tevatron.}
\label{fig:tev}
\end{figure}

It has also been asserted that there are large theoretical 
uncertainties arising from substantial QCD corrections to
$W$ production at the LHC energy.  In Fig.~\ref{fig:lhc}a.,
we present the leading order (LO) calculation and next-to-leading order (NLO)
QCD calculation~\cite{walt} of the transverse mass distribution ($m_T$) at the
LHC (14~TeV, $pp$ collider) in the region of interest for 
the extraction of the mass.    
We used the MRSA~\cite{mrsa} set of parton distribution functions, 
and imposed a charged lepton (electron or muon) rapidity cut of 1.2, 
as well as a charged lepton $p_T$ and missing transverse energy cut
of 20 GeV.  We used $m_W$ for the factorization and renormalization scales.  
No detector effects were included in our calculation.
The uncertainty due to the QCD corrections can be gauged
by considering the ratio of the NLO calculation over the LO calculation.  
This ratio is presented in Fig.~\ref{fig:lhc}b. as a function of $m_T$.
As can be seen, the corrections are not large and vary between 10\% and 20\%.  
For the extraction of $m_W$ from the data, the important consideration 
is the change in the shape of the $m_T$-distribution.  As can be seen
from Fig.~\ref{fig:lhc}b, the corrections to the shape of the 
$m_T$-distribution are at the 10\% level.  
Note that an  
increase in the charged lepton $p_T$ cut has the effect of increasing 
the size of the shape change (it basically increases the slope of the 
NLO over LO ratio), such that for the theoretical uncertainty 
is is better to keep that cut as low as possible.
For comparison, in Fig.~\ref{fig:tev} we present the same distributions as in 
Fig.~\ref{fig:lhc} for the Tevatron energy (1.8~TeV, $p \bar{p}$ collider).  
The same cuts as for the LHC were applied.  
As can be seen the corrections
are of the order of 20\% and change the shape very little.
Although the shape change due to QCD corrections 
is undoubtedly larger at the LHC 
than at the Tevatron, 
this does not appear to be a serious problem considering the size of the
corrections.  A next-to-next-to 
leading order or eventually an appropriate resummed calculation 
should be able to bring the theoretical uncertainty down to 
an acceptable level.  Although the next-to-next-to-leading order 
calculation doesn't yet exist for the $m_T$-distribution,
one may certainly imagine that it will be done before any data become 
available at the LHC.  An alternative
would be to use an observable with yet smaller QCD corrections.  
Recently~\cite{gandk}, it was pointed out that the ratio of $W$ over 
$Z$ observables (properly normalized with the mass)
are subject to smaller QCD corrections than the observables themselves.
Indeed, the corrections are similar for the $W$ and $Z$ observables and 
therefore cancel in the ratio.  The ratio of the transverse mass could be used
to measure the mass, with small theoretical uncertainty.  
Compared to the standard transverse-mass method, this method
will have a larger statistical uncertainty because it depends on the
$Z$ statistics, but a smaller systematic uncertainty because
of the use of the ratio.  Overall this ratio method might therefore
be competitive.
%

It is interesting to note that the average Bjorken-$x$ of the 
partons producing the $W$ at LHC is  $\sim 6 \times10^{-3}$ 
($\sim m_W$/energy of the collider), compared
to  $\sim 4 \times 10^{-2}$ at the Tevatron.  The uncertainty due
to the parton distributions will thus be different at the LHC and Tevatron
as different region of Bjorken-$x$ are probed.  
Considering that the uncertainty due to the parton distribution 
functions might dominate in this very high precision measurement, 
complementary
measurements at the Tevatron and LHC would be very valuable.

The production cross section at the LHC, with the cuts already mentioned
and $65 GeV\leq m_T \leq 100GeV$,  
is about $4$ times larger than at the Tevatron.
Scaling from the $9 \times 10^3$ $W$ events measured at the Tevatron with 
an integrated luminosity of $20 pb^{-1}$~\cite{cdf} (one detector), 
we then expect at the LHC $\sim 1.8 \times 10^7$ reconstructed $W$ 
events in one year at low luminosity (for $10fb^{-1}$).
(If the lepton rapidity coverage at the LHC were increased 
above the $\pm 1.2$ assumed here, a large gain in signal statistics
would be obtained, since 
the rapidity distribution is rather broad at the LHC energy.)

As a first estimate of the precision with which $m_W$ can be determined 
we have simply taken the formula
which were developed by the TeV2000 study~\cite{tev2000} 
to include the effect of $I_C$.:
\begin{eqnarray}
\Delta m_W |_{stat}& = &12.1\,{\rm GeV} \sqrt{I_C\over N} \sim 4\,{\rm MeV}
\nonumber \\
\Delta m_W |_{sys}& = &17.9\,{\rm GeV} \sqrt{I_C\over N} \sim 6\,{\rm MeV}
\end{eqnarray}
where $N$ is the total number of events.
Taken at face value these would suggest that $\Delta m_W \sim 7$~MeV could be
reached.  Systematic effects which are not yet 
important in present data could limit the attainable 
precision; but we feel that it should be
possible to measure the $W$ mass to a precision of better than
$\Delta m_W \sim 15$~MeV at the LHC.  

It is worth noting that, while we have assumed that only 
one year of operation at low luminosity is required to collect the dataset,
considerably longer would undoubtedly be required after the data are
collected, in order to understand 
the detector at the level needed to make such a precise measurement.

In conclusion, while this is a very first study of this question, we see no
serious problem with making a precise measurement of $m_W$ at the LHC 
if the accelerator is operated
at low luminosity ($10^{33}\,{\rm cm}^{-2}\,{\rm s}^{-1}$) for at least 
a year.  
The cross section is large, triggering is possible, 
lepton identification and measurement straightforward,
and the missing transverse energy should be well determined.
The QCD corrections to the transverse mass distribution
although larger than at the Tevatron, still appear reasonable.  
We imagine that a precision better than $\Delta m_W \sim 15$~MeV 
could be reached, making this measurement the world's best 
determination of the $W$ mass.
We feel that it is well worth investigating this opportunity 
in more details.

%


\begin{thebibliography}{9}

\bibitem{wolf}
see for example W.~de Boer, A.~Dabelstein, W.~Hollik, W.~M\"osle and 
U.~Schwickerath, KA-TP-18-96, hep-ph/9609209.

\bibitem{lep}
%
A.~Ballestrero {\it et al.}, in Proceedings of the Workshop on 
Physics at LEP2, G.~Altarelli,T.~Sjostrand and F.~Zwirner (eds.), 
CERN Yellow Report CERN-96-01 (1996). 
%
\bibitem{tev2000} 
`Future Electroweak Physics at the Fermilab Tevatron:
Report of the tev\_2000 Study Group,'
D.~Amidei and R.~Brock (eds.), FERMILAB-Pub-96/082, April 1996.

\bibitem{tevah}
%
The TEV33 Committee Report, Executive Summary,
http://www-theory.fnal.gov/tev33.ps.

\bibitem{cmsetrig}
`Preliminary specification of the baseline calorimeter trigger algorithms,'
J.~Varela {\it et al.}, CMS Internal Note CMS-TN/96-010, unpublished;\\
`The CMS electron/photon trigger: simulation study with CMSIM data,'
R. Nobrega and J. Varela, CMS Internal Note CMS-TN/96-021, unpublished.

\bibitem{cmsmutrig}
`CMS Muon Trigger --- Preliminary specifications of the 
baseline trigger algorithms,'
F.~Loddo {\it et al.}, CMS Internal Note CMS-TN/96-060, unpublished.

\bibitem{mrsa}
%
A.D.~Martin, R.G.~Roberts and W.J.~Stirling, Phys.~Rev.~{\bf
D50}, 6734 (1994).

\bibitem{walt}
Our calculation is based on: W.T.~Giele, E.W.N.~Glover, 
and D.A.~Kosower, Nucl.~Phys.~B403, 633 (1993). 

\bibitem{gandk}
$M_W$ Measurement at the Tevatron With High Luminosity,
W. T. Giele and S. Keller, to appear in the 
proceedings of the DPF96 Conference, Minneapolis, MN, August~10 --~15, 1996,
FERMILAB-CONF-96/307-T.

\bibitem{cdf}
F.~Abe {\it et al.}, (CDF Collaboration), Phys.~Rev.~Lett.~{\bf 75}, 11 (1995)
and Phys.~Rev.~{\bf D52}, 4784 (1995).
\end{thebibliography}
\end{document}